\documentclass[a4paper,11pt]{article}
\pdfoutput=1 

\usepackage{jheppub} 

\usepackage[T1]{fontenc} 

\def\be{\begin{equation}}
\def\ee{\end{equation}}
\newcommand{\ba}{\begin{eqnarray}}
\newcommand{\ea}{\end{eqnarray}}

\title{\rm \bf \Huge U(1) symmetric $\alpha$-attractors}

\author{\rm {\bf Yusuke Yamada}}
\affiliation{Stanford Institute for Theoretical Physics and Department of Physics, Stanford University, Stanford, CA 94305, USA}

\emailAdd{yusukeyy@stanford.edu}

\abstract{We present a class of supergravity $\alpha$-attractors with an approximate global U(1) symmetry corresponding to the axion direction. We also develop a multi-field generalization of these models and show that the $\alpha$-attractor models with U(1) symmetries have a dual description in terms of a two-form superfield coupled to a three-form superfield.}

\begin{document}
\maketitle
\flushbottom
\section{Introduction}
Inflationary cosmology becomes increasingly important from the cosmological and the theoretical physics viewpoint. The cosmic microwave background (CMB) observations have been improved, and the constraint on inflationary models became more restrictive. In particular, the CMB observations give constraints on the shape of the scalar potential if the inflation is driven by a single scalar field.

The $\alpha$-attractor models~\cite{Kallosh:2013yoa} is a class of inflationary models, which predict the spectral index of the scalar curvature perturbation $n_s=1-\frac{2}{N}$ and the tensor-to-scalar ratio $r=\frac{12\alpha}{N^2}$, where $N$ is the number of e-foldings. The value of $n_s$ nicely fits into the latest CMB observation data, independently of the value of $\alpha$ in the range $\alpha\lesssim \mathcal O(1)$ \cite{Ade:2015lrj}. The tensor-to-scalar ratio $r$ is also testable by the future experiments for $\alpha\sim \mathcal O(0.1-1)$.

The theory of  $\alpha$-attractors in supergravity can be most naturally described in terms of the hyperbolic geometry~\cite{Kallosh:2015zsa,Carrasco:2015uma,Ferrara:2016fwe,Kallosh:2017ced}, which is the moduli space geometry of the inflaton superfield,
\begin{equation}
ds^2=\frac{3\alpha (d\tau^2+d\chi^2)}{4\tau^2},
\end{equation}
where $\tau$ is the inflaton field and $\chi$ is an axion. Importantly, the observational predictions of these theories are very stable with respect to strong modifications of the inflaton potential in terms of the original geometric variables.

The hyperbolic moduli space geometry appears in extended supergravity and string/M-theory~\cite{Ferrara:2016fwe,Kallosh:2017ced}, and hence, seems well-motivated from a theoretical viewpoint. Also, such a UV theory gives seven (complex) moduli fields with $\alpha=\frac13$, and the merging mechanism can effectively lead to $\alpha=\frac13,\cdots,\frac73$, which gives the tensor-to-scalar ratio $r=\mathcal{O}(0.001-0.01)$ testable in the future CMB B-mode experiments~\cite{Ferrara:2016fwe,Kallosh:2017ced}. Such an interesting moduli space geometry has several symmetries, one of which is the shift symmetry of the inflaton. 

Most of the $\alpha$-attractor models studied so far are constructed in such a way that the inflaton shift symmetry is only slightly broken by its potential, whereas other symmetries are completely broken. Therefore only the inflaton field is light during inflation. However, it turns out that, even if the superpartner axion is light, it effectively freezes during inflation and is not harmful for successful inflation~\cite{Achucarro:2017ing}. Such light particles can result in a phenomenologically rich structure in the low energy physics  without affecting the successful inflationary predictions of the single-field $\alpha$-attractors.

Only a very narrow class of $\alpha$-attractors with such properties was known until now: T-models with $\alpha = 1/3$. In this paper, we will construct a much more broad class of $\alpha$-attractor models with global (nonlinear) U(1) symmetries in supergravity, including models with $\alpha \not = 1/3$. We assume that the U(1) symmetry is slightly broken, e.g. by instanton effects. There appears a light axion as a pseudo Nambu-Goldstone mode. We show how to realize (approximately) U(1) symmetric model in supergravity, and extend the construction to the multi-field case. 

We also discuss the dual description of the U(1) symmetric $\alpha$-attractor models by using a two-form superfield, which gives us different perspectives of the models in terms of the dual gauge (form) superfields. From such a dual viewpoint, the shift symmetry of the axion can be seen as a consequence of the gauge symmetry of the dual two-form field. As we will show, the Poincar\'e dual of the U(1) symmetric $\alpha$-attractor can be identified as the system with two- and three-form superfields. Gauging the two-form under the three-form gauge symmetry leads to a mass term of the axion in the dual scalar system as applied to natural chaotic inflation model~\cite{Kaloper:2008fb,Kaloper:2011jz,Dudas:2014pva}. Such a mass term is naturally small and as a result the light axion would appear.

This paper is organized as follows. First, in Sec.~\ref{U(1)} we briefly review the symmetry of the hyperbolic geometry and find that there is a U(1) symmetry corresponding to an axion shift symmetry. We show concrete examples of U(1) symmetric models with a single field as well as multiple moduli fields. We discuss the dual formulation of the U(1) symmetric model in Sec.~\ref{Dual}. Also we show that the deformation of the dual system naturally yields the mass of the axion. Finally, we give a summary of this paper in Sec.~\ref{summary} and briefly discuss possible applications of the U(1) symmetric models. In Appendix~\ref{app}, we discuss the possible deformation of the potential by adding a non-perturbative superpotential.
\section{U(1) symmetric $\alpha$-attractors}\label{U(1)}
\subsection{U(1) symmetry in hyperbolic geometry}
The $\alpha$-attractor models are known to be described by the hyperbolic geometry of moduli space~\cite{Kallosh:2015zsa,Carrasco:2015uma,Ferrara:2016fwe,Kallosh:2017ced}. The metric of the moduli space is given by
\begin{equation}
ds^2=3\alpha\frac{dTd\bar T}{(T+\bar{T})^2},\label{mT}
\end{equation}
or
\begin{equation}
ds^2=3\alpha\frac{dZd\bar Z}{(1-Z\bar Z)^2},\label{mZ}
\end{equation}
where $T$ and $Z$ are complex scalars and related to each other via the Cayley transformation $Z=\frac{T-1}{T+1}$. These metrics are invariant under the M\"obius transformation~\cite{Carrasco:2015uma}. For simplicity, we will take the half-plane variable $T$, which transforms under M\"obius transformation as
\begin{align}
iT\to \frac{iaT+b}{icT+d},
\end{align}
where $a,b,c,d$ are real numbers satisfying $ad-bc\neq0$. The transformation is represented by a matrix (see e.g.~\cite{Carrasco:2015uma})
\begin{equation}
\mathcal{M}=
\begin{pmatrix}
a&b\\
c&d
\end{pmatrix}.
\end{equation}
More useful representation is Iwasawa decomposition form
\begin{align} 
\mathcal{M}=&K\cdot A\cdot N\nonumber\\
=&\begin{pmatrix}\cos\mathcal\theta&-\sin\theta\\ \sin\theta&\cos\theta\end{pmatrix}\cdot\begin{pmatrix}r_1&0\\ 0&r_2\end{pmatrix}\cdot\begin{pmatrix}1&x\\0&1\end{pmatrix},
\end{align}
where
\begin{equation}
\cos\theta=\frac{a}{\sqrt{a^2+c^2}},\quad r_1=\sqrt{a^2+c^2},\quad r_2=\frac{ad-bc}{\sqrt{a^2+c^2}}, \quad x=\frac{ab+cd}{a^2+c^2}.
\end{equation}
We parametrize $T$ as
\begin{equation}
T=\exp\left({\sqrt{\frac2{3\alpha}}\varphi}\right)+i\chi,
\end{equation}
where $\varphi$ is the canonical inflaton and $\chi$ is a real scalar. Under $A$ transformation of the M\"obius group, each real scalar field transforms as
\begin{equation}
\varphi\to \varphi+\sqrt{\frac{3\alpha}{2}}\log\left(\frac ad\right),\quad \chi\to \frac ad\chi,\label{shifti}
\end{equation} 
whereas under $N$ transformation, we find
\begin{equation}
\chi\to\chi+b.\label{shifta}
\end{equation} 
These partial transformations \eqref{shifti} and \eqref{shifta} correspond to the shift symmetry of $\varphi$ and $\chi$, respectively. The shift symmetry of $\varphi$ is responsible for successful inflation and slightly broken by scalar potential of $\varphi$. 

Under these transformations, the K\"ahler metric is invariant, whereas the K\"ahler potential transforms,
\begin{equation}
K=-3\alpha\log(T+\bar{T})\to K-3\log(\frac{a}{d}).
\end{equation}
Therefore, we find that the K\"ahler potential is variant under $A$ transformation. Although the $A$-invariant K\"ahler potential exists~\cite{Carrasco:2015uma}, we consider the models with a light axionic direction where $N$-invariance is responsible. Hence, in the following, we will construct models with $N$-invariant K\"ahler potential $K=-3\alpha\log(T+\bar{T})$.

One may think of the shift symmetry of $\chi$ generated by $N$ as nonlinearly realized global U(1) symmetry. Such a U(1) symmetry is known to appear e.g. in Green-Schwarz mechanism~\cite{Green:1984sg,LopesCardoso:1991ifk}. We stress that this U(1) symmetry is independent of the shift symmetry of the inflaton $\varphi$, and therefore, it is possible to construct inflationary models which preserve the U(1) symmetry. 

We find the following U(1) symmetry for the metric \eqref{mZ},
\begin{align}
Z\to Ze^{i\eta},\label{zut}
\end{align}
where $\eta$ is a real constant. For a disk variable $Z$, we use the parametrization,
\begin{equation}
Z=e^{i\theta}\tanh\left(\frac{\varphi}{\sqrt{6\alpha}}\right),
\end{equation}
where $\theta$ is a real scalar and $\varphi$ is (canonical) inflaton field. The U(1) transformation~\eqref{zut} is the shift of $\theta$,
\begin{equation}
\theta\to\theta+\eta.
\end{equation}
If this U(1) symmetry is broken spontaneously, $\theta$ becomes a Nambu-Goldstone boson of this symmetry.
Thus, for both $T$ and $Z$-variables, we find (nonlinear) global U(1) symmetries, and the partners of the inflaton fields become Nambu-Goldstone modes of these U(1) symmetries if they are spontaneously broken. We note that, in our models discussed below, the U(1) symmetry is spontaneously broken during inflation, and therefore, the domain wall and cosmic string formation after inflation do not occur. 
\subsection{single-disk model}
We consider U(1) symmetric $\alpha$-attractors in supergravity framework. If we construct inflation models with superpotential terms, it is difficult to realize various potential for inflaton while preserving the U(1) symmetry. The U(1) symmetry tends to be broken at inflation scale in such a case. The $\overline{\text D3}$-induced geometric inflation~\cite{McDonough:2016der,Kallosh:2017wnt} is useful to construct such U(1) symmetric systems in a simple way since the inflaton potential originates from a K\"ahler coupling between the inflaton and a nilpotent superfield $S$~\cite{Rocek:1978nb,Lindstrom:1979kq,Casalbuoni:1988xh,Komargodski:2009rz,Kuzenko:2010ef}.

For a half-plane variable $T$, we consider the following K\"ahler and super-potential\footnote{We are able to rewrite them with K\"ahler invariant function $\mathcal G$ through the relation
$\mathcal G = K+\log |W|^2$.}
\begin{align}
K=&-3\alpha\log(T+\bar T)+G_{S\bar S}S\bar S,\\
W=&W_0(1+S),
\end{align}
where $S$ is a nilpotent superfield and $W_0$ is a constant. Here, we choose $G_{S\bar S}$ as
\begin{align}
G_{S\bar S}=\frac{|W_0|^2}{(T+\bar T)^{3\alpha}V_0(T,\bar T)+3|W_0|^2(1-\alpha)},
\end{align}
where $V_0(T,\bar{T})$ is an arbitrary function of $T$ and $\bar T$. Using the standard supergravity formula with $S=0$ projection, one finds the scalar potential
\begin{equation}
V=V_0(T,\bar T).
\end{equation}
Choosing $V_0(T,\bar{T})=f(T+\bar T)$ for an arbitrary function $f$, the system becomes manifestly invariant under the nonlinear U(1) transformation $T\to T+i b$. In such case, the potential does not have $\chi$ dependence, and therefore $\chi$ is completely massless. One may introduce small corrections $\Delta V_0(T,\bar{T})$, which break the U(1) symmetry and gives a small mass to the axion $\chi={\rm Im}T$. Note that there is a constraint on the possible value of $\alpha$: since the potential $V_0$ is almost vanishing at the vacuum, we have to require $\alpha<1$ for $G_{S\bar S}$ to be positive definite.\footnote{$\alpha=1$ is allowed from this constraint and $G_{S\bar S}$ becomes very simple. However, in such a case, $G_{S\bar S}$ at $V_0\sim0$ becomes extremely large, which might lead to a strong coupling after rescaling the nilpotent superfield to make it canonical. A possible way to realize $\alpha=1$ is e.g. to introduce additional source of negative contribution to the energy. Then, $V_0\sim0$ at the minimum is not required, and $G_{S\bar{S}}$ becomes non-singular. However, we also note that introducing additional negative contribution is not simple by the following reason: The usual negative term in F-term potential $-3e^K|W|^2$ is canceled by the property of the no-scale structure. Hence, the nontrivial modification of the model is necessary for the realization of $\alpha=1$ model. We do not treat such case in the following.} It is possible to introduce an axion mass via superpotential. We have shown an example of such case in appendix~\ref{app}.

For a disk variable $Z$, we consider the following system
\begin{align}
K=&-3\alpha\log(1-Z\bar Z)+G_{S\bar S}S\bar S,\\
W=&W_0(1+S),\\
G_{S\bar S}=&\frac{|W_0|^2}{(1-Z\bar Z)^{3\alpha}\tilde{V}_0(Z,\bar Z)+3|W_0|^2(1-\alpha Z\bar Z)},
\end{align}
where $\tilde{V}_0(Z,\bar Z)$ is a real function of $Z$ and $\bar{Z}$. The scalar potential in this system is
\begin{align}
V=\tilde{V}_0(Z,\bar Z).
\end{align}
As in the case of the half-plane variable $T$, for $\tilde{V}_0=g(Z\bar Z)$ with an arbitrary $g$, the model becomes manifestly invariant under the transformation~\eqref{zut} or equivalently, the potential is independent of $\theta$. The small deviation from the exact U(1) symmetric potential gives rise to the small potential for the axion $\theta={\rm Arg}Z$. Note also that $\alpha<1$ is required for the consistency $G_{S\bar S}>0$ at the vacuum $\tilde{V}_0\sim 0$. We note that the model in \cite{Achucarro:2017ing} corresponds to our model with $\alpha=\frac13$. We stress that  the models shown above are relatively simple construction of U(1) symmetric $\alpha$-attractors, but there would be different constructions preserving the U(1) symmetries. Also it is possible to construct models with an unconstrained superfield instead of a nilpotent superfield $S$, as we will show in Sec.~\ref{linear}.
\subsection{multi-disk model}
We develop the multiple-moduli generalization of the U(1) symmetric $\alpha$-attractors, which is straightforward and useful for model buildings.

Let us consider the system with $n$ half-plane variables $T_i$,
\begin{align}
K=&\sum_i^n-3\alpha_i\log(T_i+\bar T_i)+G_{S\bar S}S\bar S,\\
W=&W_0(1+S).
\end{align}
Here, we do not need to restrict the number $n$. Note that, however, maximal supersymmetry and superstring/M-theory realize $n=7$ disk moduli fields with $\alpha_i=\frac13$ in their four-dimensional $N=1$ reduction~\cite{Ferrara:2016fwe}. We choose $G_{S\bar S}$ as
\begin{align}
G_{S\bar S}=\frac{|W_0|^2}{V_0\prod_i^n(T_i+\bar{T_i})^{3\alpha_i}+3|W_0|^2(1-\sum_i^n \alpha_i)},
\end{align}
where $V_0=V_0(T_1,\bar T_1,\cdots, T_n,\bar T_n)$ is a real function. The $U(1)^n$ symmetry is realized when $V_0=f(T_1+\bar T_1,\cdots, T_n+\bar T_n)$. As in the models with a single field, we find a constraint on the choice of model, 
\begin{equation}
\sum_i^n\alpha_i<1,\label{multiC}
\end{equation}
so that $G_{S\bar S}>0$.

In the similar way, we extend the disk-variable model as
\begin{align}
K=&\sum_i^n -3\alpha_i\log (1-Z_i\bar Z_i)+G_{S\bar S}S\bar S,\\
W=&W_0(1+S).
\end{align}
Here
\begin{equation}
G_{S\bar S}=\frac{|W_0|^2}{\tilde{V}_0\prod_i^n (1-Z_i\bar Z_i)^{3\alpha_i}+3|W_0|^2(1-\sum \alpha_i Z_i\bar Z_i)},
\end{equation}
where $\tilde{V}_0$ is a real function of $Z_1,\bar Z_1,\cdots,Z_n,\bar Z_n$. This system has global $U(1)^n$ manifestly. Note that the constraint on this model for $G_{S\bar S}>0$ is the same as \eqref{multiC} since each $Z_i$ satisfies $Z_i<1$. The scalar potential of this system is given by
\begin{equation}
V=\tilde V_0(Z_1\bar Z_1,\cdots,Z_n\bar Z_n).
\end{equation}

The presence of multiple disk fields can realize the various values of the tensor-to-scalar ratio $r$ due to the merger of $\alpha$-attractors~\cite{Kallosh:2017ced,Kallosh:2017wnt,Kallosh:2017wku}. It would be a way to realize various values of $r$ against the restriction $\sum_i\alpha_i<1$.
\subsection{A model with an unconstrained stabilizer}\label{linear}
So far, we have used a nilpotent superfield in our model buildings. Here, let us construct models with an unconstrained superfield instead of a nilpotent superfield. For simplicity, we consider a model with a single half-plane $T$ with $\alpha=\frac23$, but in the similar way one can make e.g. models with multiple-disk moduli. Let us consider the following system
\begin{align}
K=&-2\log(2\tau)+G_{X\bar X}X\bar X-\zeta (X\bar{X})^2,\\
W=&W_0(1+X),
\end{align}
where $\zeta$ is a positive constant, $\tau=\frac12(T+\bar{T})$ and
\begin{equation}
G_{S\bar S}=\frac{|W_0|^2}{\tau^2V_0(\tau)+|W_0|^2}.
\end{equation}
Here, $X$ is not a nilpotent superfield but an unconstrained superfield. The quartic term of $X$ in K\"ahler potential is important for stabilization of $X=\rho+i\psi$. One can expand the potential around $\rho=\psi=0$ up to quadratic terms of them and find the effective minimum for $\rho$ and $\psi$
\begin{equation}
\rho=\frac{W_0(2e^{-2\phi}V_0+e^{-3\phi}V'_0)}{\zeta (W_0+4e^{-2\phi}V_0)^3},\quad \psi=0,
\end{equation}
where the prime denotes the derivative with respect to $\tau$ and we have used $\tau=e^{-\phi}$ where $\phi$ is the canonical inflaton field. We find that the expectation value of $\rho$ becomes exponentially suppressed. Also, it is further suppressed when $W_0\sim m_{3/2}>\sqrt{ V_0}\sim H_{\rm inf}$, which is necessary to avoid the gravitino problem~\cite{Hasegawa:2017nks}. Therefore, the correction to potential originating from the VEV of $X$ is negligibly small. The effective potential including the leading order correction from $X$ is given by
\begin{equation}
V_{\rm eff}=V_0\left(1-\frac{4e^{-6\phi}\delta^4\left(1+e^{-\phi}\frac{V'_0}{2V_0}\right)^2}{\zeta(1+4e^{-2\phi}\delta)^4}\right).
\end{equation}
The corrections are suppressed by $e^{-\phi}$ factors as well as $\delta=V_0/W_0^2$ and hence the inflationary prediction is the same as the models with a nilpotent superfield. Thus, we have shown that the U(1) symmetric $\alpha$-attractors can be realized with an unconstrained superfield. Note also that $X$ behaves as a heavy Polonyi field at the vacuum $V_0\sim0$, and supersymmetry is not restored unlike models with stabilizer fields.

\begin{figure}[h!]
\begin{center}
\includegraphics[scale=0.3]{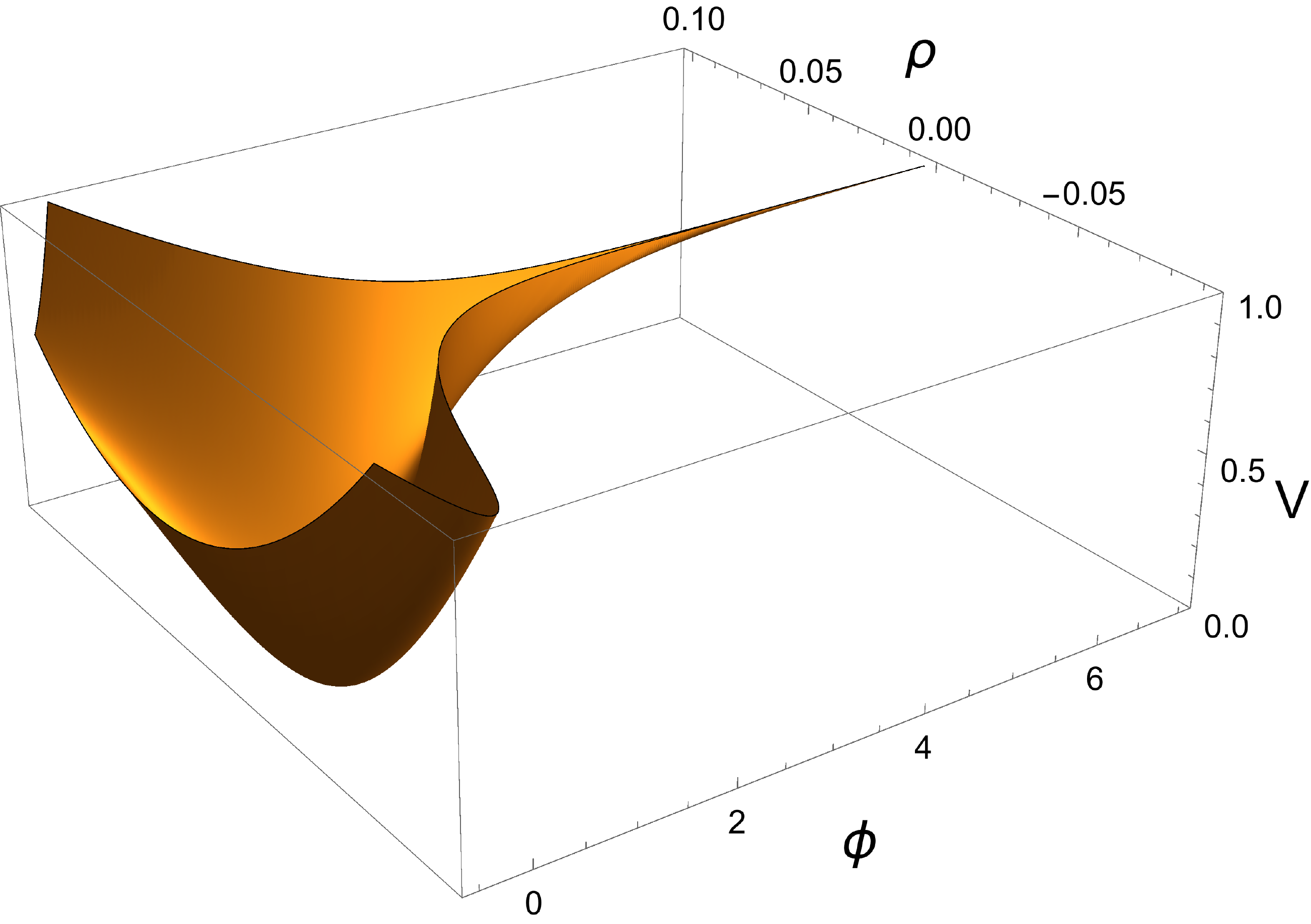}
\end{center}
\caption{\footnotesize The scalar potential for $\phi$ and $\rho$ with $V_0=M^2(1-\tau)^2$. }
\label{plot}
\end{figure}
As a simple example, we show the potential for $\rho$ and $\phi$ with $V_0=M^2(1-\tau)^2$ in Fig.~\ref{plot}. Here we have taken the parameters $W_0=10M^2, \ \zeta=1,\ M=1$ and $\psi$ is set at $\psi=0$. Inflation takes place along the very narrow line $\rho\sim0$, and this shows the stability of the inflationary trajectory.

\section{Dual description of U(1) symmetric $\alpha$-attractor}\label{Dual}
In this section, we will discuss the dual formulation of the U(1) symmetric models, which might be useful e.g. to consider the UV completion of our models. It is well known that in four-dimension, the Poincar\'e dual of an axionic field is the two-form field $B_{\mu\nu}$, of which the field strength is the three-form $H_{\mu\nu\rho}=\partial_{[\mu}B_{\nu\rho]}$. There is a supersymmetric version of the Poincar\'e dual, which converts a chiral superfield to a real linear superfield $L$. The linear-chiral duality is used e.g. in the low energy effective theory of string theory~\cite{Burgess:1995kp,Burgess:1995aa,Binetruy:1995hq}. A linear superfield $L$ is a real superfield satisfying a constraint $D^2L=\bar{D}^2L=0$. It consists of a real scalar $C$, a spinor $\zeta$ and a 3-form $H_{\mu\nu\rho}$. In a flat superspace, 
\begin{equation}
L=C+\theta\zeta+\bar\theta \bar\zeta+\frac12\theta\sigma_a\bar\theta \epsilon^{abcd}H_{bcd}-\frac{i}{2}\theta^2\bar\theta\bar\sigma^a\partial_a\zeta-\frac{i}{2}\bar{\theta}^2\theta\sigma^a\partial_a\bar\zeta-\frac14 \theta^2\bar{\theta}^2\Box C.
\end{equation}

We consider the dual description of the models with a single half-plane variable $T$ with $\alpha<1$.\footnote{The Poincar\`e dual of the disk-variable $Z$ is not known and we will not discuss the dual description of the disk models in this paper.} The duality with multiple half-plane variables is possible in the same way as the case with a single disk. We will consider the U(1) symmetric $\alpha$-attractor with an unconstrained superfield $X$ discussed in Sec.~\ref{linear}, and the generalization to the model with a nilpotent superfield is also possible. Using the superconformal formalism~\cite{Kaku:1978nz,Kaku:1978ea,Townsend:1979ki,Kugo:1982cu,Kugo:1983mv},\footnote{See \cite{Freedman:2012zz} for review.} the action is given by
\begin{align}
&\left[-\frac32 S_0\bar{S}_0e^{-\frac K3}\right]_D+\left[S_0^3W_0(X+1)\right]_F\nonumber\\
=&\left[-\frac32 S_0\bar{S}_0\tau^\alpha e^{\tilde{K}}\right]_D+\left[S_0^3W_0(X+1)\right]_F,\label{original}
\end{align}
where $S_0$ is a conformal compensator chiral superfield, $\tau=(T+\bar{T})$ and 
\begin{equation}
\tilde{K}(\tau,X\bar X)=-\frac13\left(G_{X\bar{X}}(\tau)X\bar{X}-\zeta f(\tau)(X\bar X)^2\right).
\end{equation}
$[\cdots]_{F,D}$ are superconformal version of superspace integrals $\int d^2\theta$ and $\int d^4\theta$, respectively. We rewrite the Lagrangian~\eqref{original} by using a linear superfield $L$ and a real general superfield $V$ as
\begin{equation}
\left[-\frac32 S_0\bar{S}_0V^\alpha e^{\tilde{K}(U,X\bar{X})}\right]_D+\left[S_0^3W_0(S+1)\right]_F+[LV]_D.\label{dual1}
\end{equation}
The variation of $L$ gives
\begin{equation}
V=\tau
\end{equation}
since, by the definition of $L$, $[L(\Phi+\bar\Phi)]_{D}=0$ for any chiral superfield $\Phi$ (see e.g.~\cite{Kugo:1982cu}). Substituting the solution, the action~\eqref{dual1} reproduces the original one~\eqref{original}. If instead we vary the real general superfield $V$, we obtain\footnote{For $\alpha=1$, this equation becomes $e^{\tilde{K}}+V\partial_V\tilde{K}=\frac{2L}{3S_0\bar{S}_0}$. We formally rewrite this equation as $V=\left(-e^{\tilde{K}}+\frac{2L}{3S_0\bar{S}_0}\right)(\partial_V\tilde{K})^{-1}$. Since $\partial_V\tilde{K}$ is proportional to $X\bar{X}$, this equation becomes singular at $X=0$, which is always the case if $X$ is a nilpotent superfield. }
\begin{equation}
V^{\alpha}e^{\tilde{K}}\left(\frac{1}{V}+\frac{1}{\alpha}\partial_V\tilde{K}\right)=\frac{2L}{3\alpha S_0\bar{S}_0}.\label{eom}
\end{equation}
One can in principle solve this equation with respect to $V$ and then $V$ becomes a function of $L$ and $X\bar X$, i.e. $V=V(\frac{L}{S_0\bar{S}_0},X\bar X)$, although it is difficult to solve the equation explicitly. We formally write the following dual action
\begin{equation}
\left[-\frac32\alpha S_0\bar{S_0}F\left(\frac{L}{S_0\bar{S}_0},X\bar{X}\right)\right]_D+[S_0^3W_0(X+1)]_F,\label{dual2}
\end{equation}
where
\begin{equation}
F\left(\frac{L}{S_0\bar{S}_0},X\bar{X}\right)=\left(\frac1\alpha V^\alpha e^{\tilde{K}(V,X\bar{X})}-\frac{2L}{3\alpha S_0\bar{S}_0}V\right)\Biggl|_{V=V\left({L}/{S_0\bar{S}_0},X\bar X\right)}.\label{defF}
\end{equation}
One can always perform the inverse procedure and find the original $\alpha$-attractor models. Thus, it is shown that the U(1) symmetric $\alpha$-attractor has a dual description with a two-form gauge field.

An interesting interpretation is to identify $X$ as a field strength of a three-form superfield~\cite{Gates:1980ay}, which includes a 4-form field strength $F_{\mu\nu\rho\sigma}=\partial_{[\mu}C_{\nu\rho\sigma]}$. Such a three-form superfield was also studied in supergravity~\cite{Binetruy:1996xw,Ovrut:1997ur,Binetruy:2000zx,Groh:2012tf}. Then, $X$ is defined as $X=-\frac14\bar{D}^2U$ where $U(=\bar U)$ is a three-form superfield~\cite{Gates:1980ay,Dudas:2014pva},
\begin{align}
U=&B+i(\theta\psi-\bar\theta\bar\psi)+\theta^2\bar x+\bar{\theta}^2x+\frac13\theta \sigma^d\bar\theta\epsilon_{abcd}C^{bcd}+\theta^2\bar\theta\left(\sqrt2\bar\lambda+\frac12\bar\sigma^a\partial_a\psi\right)\nonumber\\
&+\bar{\theta}^2\theta\left(\sqrt2\lambda-\frac12\sigma^a\partial_a\bar\psi\right)+\theta^2\bar{\theta}^2\left(D-\frac14\Box B\right).
\end{align}
Here $B$ and $D$ are real scalars, $M$ is a complex scalar, $\psi$ and $\lambda$ are Weyl spinors. In terms of these components, $X$ is expressed as
 \begin{equation}
X=x+\sqrt2\theta\lambda+\theta^2 \left(D+\frac{i}{4!}\epsilon^{abcd}F_{abcd}\right).
\end{equation}
From such a viewpoint, the system~\eqref{dual2} can be understood as the two-and three-form coupled system. However, there is no gauge symmetry connecting them in the system~\eqref{dual2}. One may introduce the following gauge symmetry, under which they transform as $C_{\mu\nu\rho}\to C_{\mu\nu\rho}+\partial_{[\mu}A_{\nu\rho]}$ and $H_{\mu\nu\rho}\to H_{\mu\nu\rho}+g\partial_{[\mu}A_{\nu\rho]}$. The supersymmetric version of the gauge transformation is introduced by modifying the linear multiplet condition as
\begin{equation}
\bar{D}^2L=\frac12mX,\label{mod}
\end{equation}
where $m$ is a constant corresponding to the gauge coupling. This constraint is invariant under the Stueckelberg-like transformation $U\to U+V$ and $L\to L+\frac12m V$, where $V$ is a real superfield. The modified linear superfield appears in the context of the Abelian tensor hierarchy~\cite{Becker:2016xgv,Becker:2016rku,Aoki:2016rfz,Yokokura:2016xcf,Becker:2017njd}, which is related to the gauge structure in higher-dimensional theories, such as eleven-dimensional supergravity~\cite{Becker:2017zwe}. For $m\to0$, the original linear multiplet condition is restored. 

\begin{figure}[h!]
\begin{center}
\includegraphics[scale=0.3]{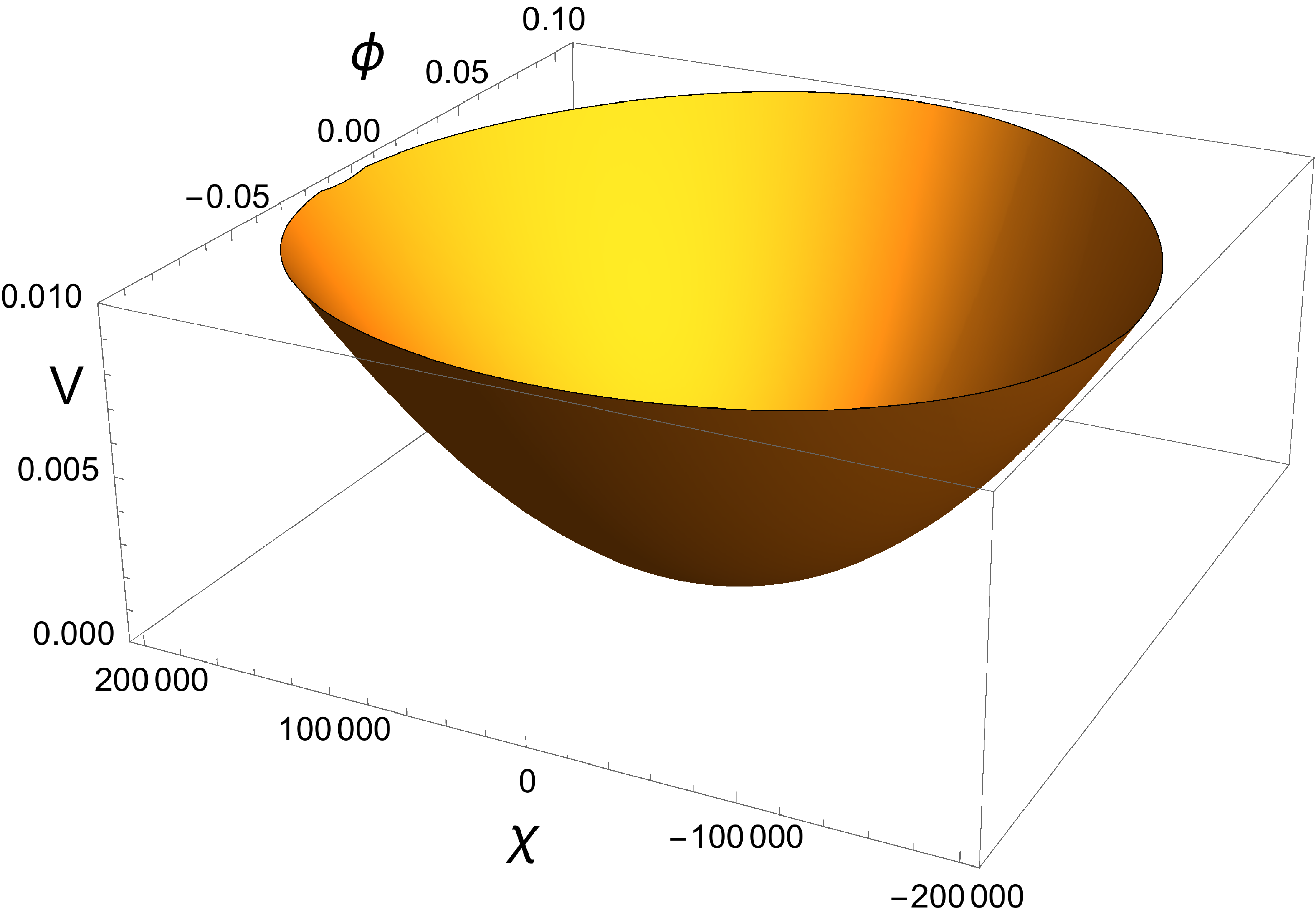}
\end{center}
\caption{\footnotesize The scalar potential for $\phi$ and $\chi$ with wide field range for $\chi$ and narrow for $\phi$. }
\label{plot2}
\end{figure}

\begin{figure}[h!]
\begin{center}
\includegraphics[scale=0.3]{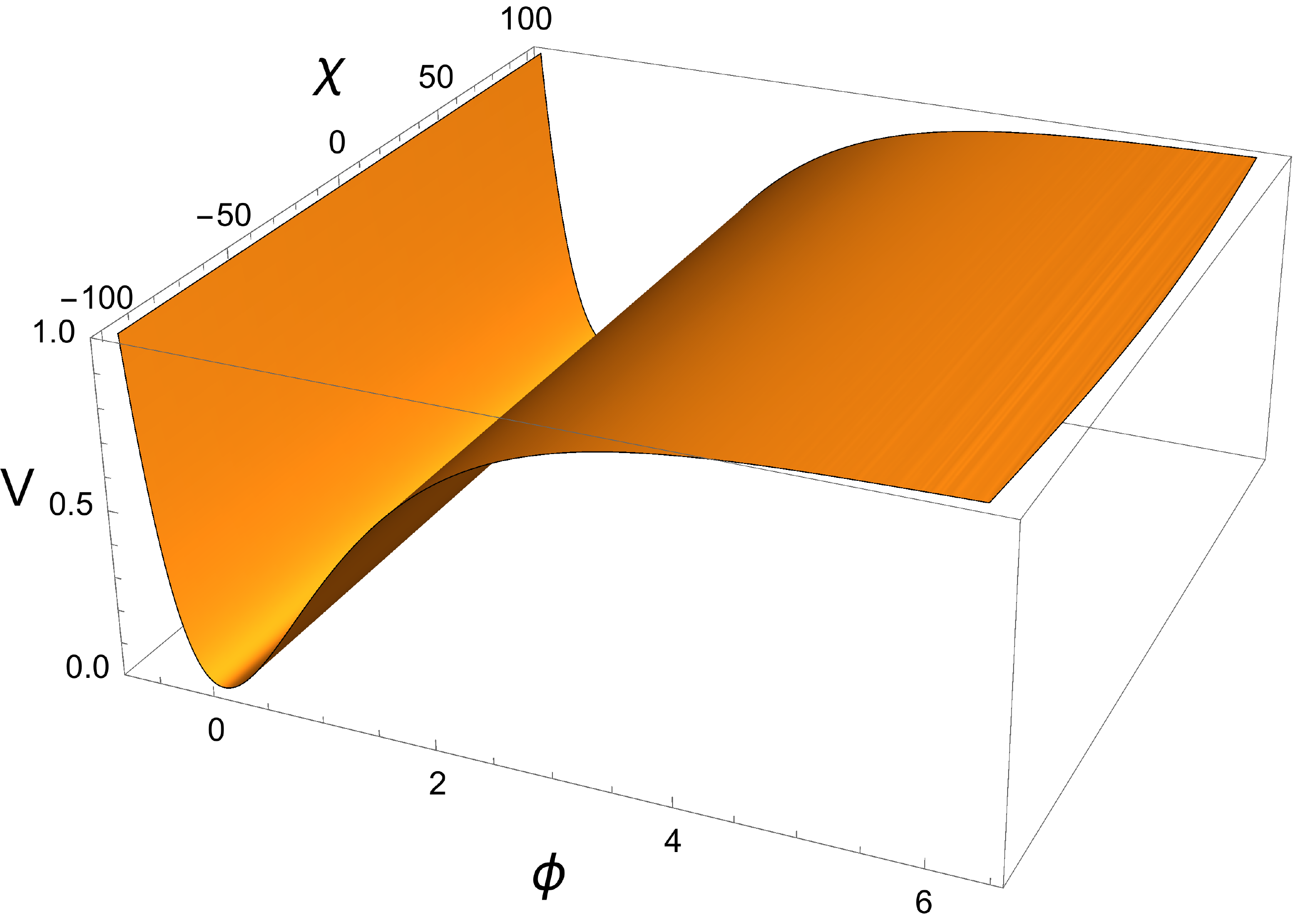}
\end{center}
\caption{\footnotesize The scalar potential for $\phi$ and $\chi$ with for field ranges different from that of Fig.~\ref{plot2}. }
\label{plot3}
\end{figure}

Let us plug into the system~\eqref{dual2} the modified linear superfield satisfying~\eqref{mod} and take the duality transformation. We modify the action~\eqref{dual2} as
\begin{equation}
\left[-\frac32\alpha S_0\bar{S_0}F\left(\frac{\tilde{L}}{S_0\bar{S}_0},X\bar{X}\right)\right]_D+[S_0^3W_0(X+1)]_F+[2S_0^3T(\Sigma(\tilde{L})-\frac12mX)]_F,
\end{equation} 
where $\tilde{L}$ is a real general superfield and $\Phi$ is a Lagrange multiplier chiral superfield. $\Sigma(\cdot)$ is a chiral projection operator corresponding to $\bar{D}^2$ in a flat superspace. The equation of motion of $T$ reads \eqref{mod} on $\tilde{L}$.  Using identity $[2S_0^3T(\Sigma(\tilde{L})]_F=[-S_0\bar{S}_0(T+\bar T)\frac{\tilde L}{S_0\bar{S}_0}]_D$, we find
\begin{equation}
\left[-\frac32\alpha S_0\bar{S_0}F\left(\frac{\tilde{L}}{S_0\bar{S}_0},X\bar{X}\right)-\tilde{L}(T+\bar{T})\right]_D+[S_0^3\left(W_0(X+1)-mTX\right)]_F.
\end{equation}
Instead of varying $T$, we consider the variation of $\tilde{L}$, which yields
\begin{equation}
V=T+\bar T,
\end{equation}
where we have used \eqref{eom} and \eqref{defF}. Thus we find the dual system with the modified linear superfield as
\begin{equation}
\left[-\frac32 S_0\bar{S}_0e^{-\frac K3}\right]_D+\left[S_0^3\left(W_0(X+1)-mTX\right)\right]_F,
\end{equation}
with
\begin{equation}
K=-3\alpha\log(T+\bar{T})+G_{X\bar X}X\bar X-\zeta f(X\bar X)^2.
\end{equation}
 Due to the gauge coupling between the two- and three-forms, there appears a new interaction, $W=-mTX$, which depends on the "gauge coupling" $m$. Since this superpotential breaks a U(1) symmetry, the axion acquires a mass from this coupling. Such a mechanism is used in the context of the natural chaotic inflation~~\cite{Kaloper:2008fb,Kaloper:2011jz,Dudas:2014pva}. In the modified U(1) $\alpha$-attractor model, for the coupling $m$ much smaller than the inflation scale, the inflation takes place as the undeformed case while the axion acquires a tiny mass term. We show an example with
 \begin{equation}
 V_0=M^2\left(1-\frac12(T+\bar T)\right)^2,\ f=\zeta=M=1,\ m=10^{-6}, \alpha=\frac23
 \end{equation}
 where $T=e^{-\phi}+i\chi$.
 In this case, the additional superpotential does not change the inflaton potential while it gives small mass term to the axion $\chi$ as shown in Figs.~\ref{plot2} and \ref{plot3}.\footnote{Note that this example does not realize the small cosmological constant $\Lambda^4=\mathcal{O}(10^{-120})$ consistent with observations. For a realistic model building, one needs e.g. to shift $V_0$ by introducing additional parameter such that the potential minimum realizes a tiny cosmological constant.}

 In this dual formulation, although the hyperbolic moduli space geometry is not manifest, the shift symmetry of the axion can be seen as the consequence of the two-form gauge symmetry. Also one can introduce the axion mass as the coupling between dual two- and three-forms. Note also that we have identified only $X$ as a three-form superfield. However, it is also possible to think of the chiral compensator $S_0$ as a three-form multiplet as discussed in~\cite{Ovrut:1997ur,Farakos:2017jme}. Such a description might relate the models e.g. to the flux compactification in higher-dimensional supergravity.
\section{Summary and discussion}\label{summary}
We have constructed a class of $\alpha$-attractor models which has global U(1) symmetry. Such a symmetry is a part of the M\"obius transformation, under which the hyperbolic geometry is invariant. The geometric model building studied in \cite{McDonough:2016der,Kallosh:2017wnt} is suitable since it does not require superpotential of inflaton. The multi-field extension is straightforwardly done, which is useful to realize various values of $\alpha$ effectively. We have also discussed the model without a nilpotent superfield, which shows the same behavior as the models with a nilpotent superfield. 

We have also discussed the dual formulation of the U(1) symmetric $\alpha$-attractors, in which the shift symmetry of the axion is interpreted as a consequence of the dual two-form field. Also, if we identify the SUSY breaking field as a three-form superfield, whose F-term includes the four-form field strength $F_{\mu\nu\rho\sigma}$. From the dual viewpoint, the U(1) symmetric $\alpha$-attractor model is understood as the system where supersymmetric two- and three-forms are coupled to each other without any gauge coupling between them. As we have discussed in Sec.~\ref{Dual}, one can naturally introduce the coupling between the two-form and three-form and find a mass term for the axion field. It would be interesting to study the relation to the tensor hierarchy, which would originate from more fundamental theory.

Finally, we comment on the possible applications of the U(1) symmetric models. As shown in~\cite{Achucarro:2017ing}, the axion freezes on its potential during inflation due to an exponentially large kinetic coefficient. Then, the axion field does not acquire large quantum fluctuations despite its lightness. Therefore, the axion in the hyperbolic geometry can avoid the isocurvature perturbations problem~\cite{Ema:2016ops,ours}. A quintessence model would also be interesting direction as the application of the light axion. The application of $\alpha$-attractors to quintessence models is studied in~\cite{Linder:2015qxa,Dimopoulos:2017zvq,Mishra:2017ehw,Dimopoulos:2017tud,Akrami:2017cir}. In the presence of an additional light axion field, the effective equation of state would show a different behavior. Since we are able to introduce multiple disk moduli fields, it might be possible to realize dark matter and dark energy simultaneously within this model. Another application is the hyperinflation~\cite{Brown:2017osf,Mizuno:2017idt}, in which the "centrifugal force" to the inflaton from a massless axion leads to inflation. Our framework may help to construct a supersymmetric version of hyperinflation, which requires hyperbolic geometry with the U(1) invariant potential. We hope to return to investigation of these issues in future work.
\section*{Acknowledgement}
I am grateful to Renata Kallosh and Andrei Linde for useful discussion and comments. I am also grateful to Shuntaro Aoki, Tetsutaro Higaki, and Ryo Yokokura for the collaboration in the work~\cite{Aoki:2016rfz} and useful discussion related to the topic in Sec.~3. This work is supported by SITP and by the US National Science Foundation grant PHY-1720397.
\appendix
\section{Possible corrections}\label{app}
In this section, we show an example of U(1) symmetric model including a non-perturbative superpotential term. For concreteness, let us take the E-model in the following discussion. We firstly consider a non-perturbative superpotential correction, which appears e.g. from gaugino condensation,
\begin{equation}
\Delta W=A e^{-aT},
\end{equation}
where $A$ and $a$ are real parameters. Such contributions to superpotential break the U(1) symmetry to a discrete one, and give mass of the axion direction $\chi={\rm Im}T$. In general, inflation takes place $T<1$ or $T>1$, depending on the potential we use. For example, we can realize a potential $V=\left(1-e^{-\sqrt{\frac{2}{3\alpha}\phi}}\right)^2$ by choosing $V_0$ as
\begin{equation}
V_0=m^2\left(1-\frac12(T+\bar{T})\right)^2,
\end{equation}
where inflationary plateau region appears in ${\rm Re}T\ll1$. In this case a natural parametrization is ${\rm Re}T=e^{-\sqrt{\frac{2}{3\alpha}\phi}}$.
We also find another possibility: we could choose
\begin{equation}
V_0=m^2\left(1-\frac{2}{(T+\bar{T})}\right)^2,
\end{equation}
and then inflation takes place for ${\rm Re}T\gg1$. With this potential, it is natural to use the parametrization ${\rm Re}T=e^{\sqrt{\frac{2}{3\alpha}\phi}}$. For both cases, the potential becomes $V=\left(1-e^{-\sqrt{\frac{2}{3\alpha}\phi}}\right)^2$.

For the latter case in which inflation takes place for $T\gg1$, the non-perturbative correction $\Delta W$ does not affect inflation as long as $AW_0\ll m^2$.\footnote{Since we are assuming that $W_0$ is comparable with inflation scale, the leading contribution comes from the cross term of superpotential terms, $\sim Ae^{-aT}W_0$.} The additional contribution to the potential becomes important only near the vacuum $\phi\sim0$, and there the axion acquires a small mass from the non-perturbative correction $m_\chi^2\sim AW_0\ll m^2$.

For the former case where $T\ll1$ during inflation, due to the overall factor in F-term scalar potential, $e^K\sim \frac{1}{(T+\bar{T})^{3\alpha}}$, the small non-perturbative effect $\Delta W$ can be enhanced for $T+\bar{T}\ll1$. Indeed, the non-perturbative term gives rise to additional contributions
\begin{equation}
\Delta V\sim e^{3\alpha\sqrt{\frac{2}{3\alpha}}\phi}AW_0e^{e^{-\sqrt{\frac{2}{3\alpha}}\phi}}\cos(a\chi)+\cdots,
\end{equation}
where we have parametrized $T=e^{-\sqrt{\frac{2}{3\alpha}}\phi}+i\chi$ and we have neglected $\mathcal{O}(1)$ coefficient. The ellipses denote subdominant terms. Such contribution can be negative at $a\chi=(2n+1)\pi$ ($n\in \mathbb{Z}$), and will be the dominant term in potential for $\phi\gg1$. Therefore, adding the non-perturbative correction $\Delta W$ leads to an unacceptable infinitely negative vacuum. However, this problematic situation is easily circumvented by considering further corrections to the potential. As an example, we add the following correction
\begin{equation} 
\Delta V_0 =\frac{\delta}{(T+\bar{T})^{3\alpha+1}},
\end{equation}
where $\delta$ is a constant. In order to avoid negative infinity for $T+\bar{T}\ll1$, the parameter $\delta$ needs to be positive. As long as $\delta>0$, the potential for $T+\bar{T}\ll1$ is dominated by this term and the infinite negative energy vacuum does not show up. A problem associated with this deformation is that potential for sufficiently small $T+\bar{T}$ no longer possesses the plateau because the steep potential term $\Delta V_0$ becomes dominant. The value of $\phi$ at which potential bending takes place depends on the value of $\delta$, which can be determined by the requirement $V\sim 0$ at the potential minimum. As an illustration, we show an example of the deformed model:
\begin{align}
K=&-3\alpha\log(T+\bar T)+G_{S\bar S}S\bar S,\\
W=&W_0(1+S)+Ae^{-aT}.
\end{align}
Here, we choose $G_{S\bar S}$ as
\begin{align}
G_{S\bar S}=\frac{|W_0|^2}{(T+\bar T)^{3\alpha}V_0+3|W_0|^2(1-\alpha)},
\end{align}
and
\begin{equation}
V_0=m^2\left(1-\frac12(T+\bar{T})\right)^2+\frac{\delta}{(T+\bar{T})^{3\alpha+1}}.
\end{equation}
\begin{figure}[t]
\begin{center}
\includegraphics[scale=0.5]{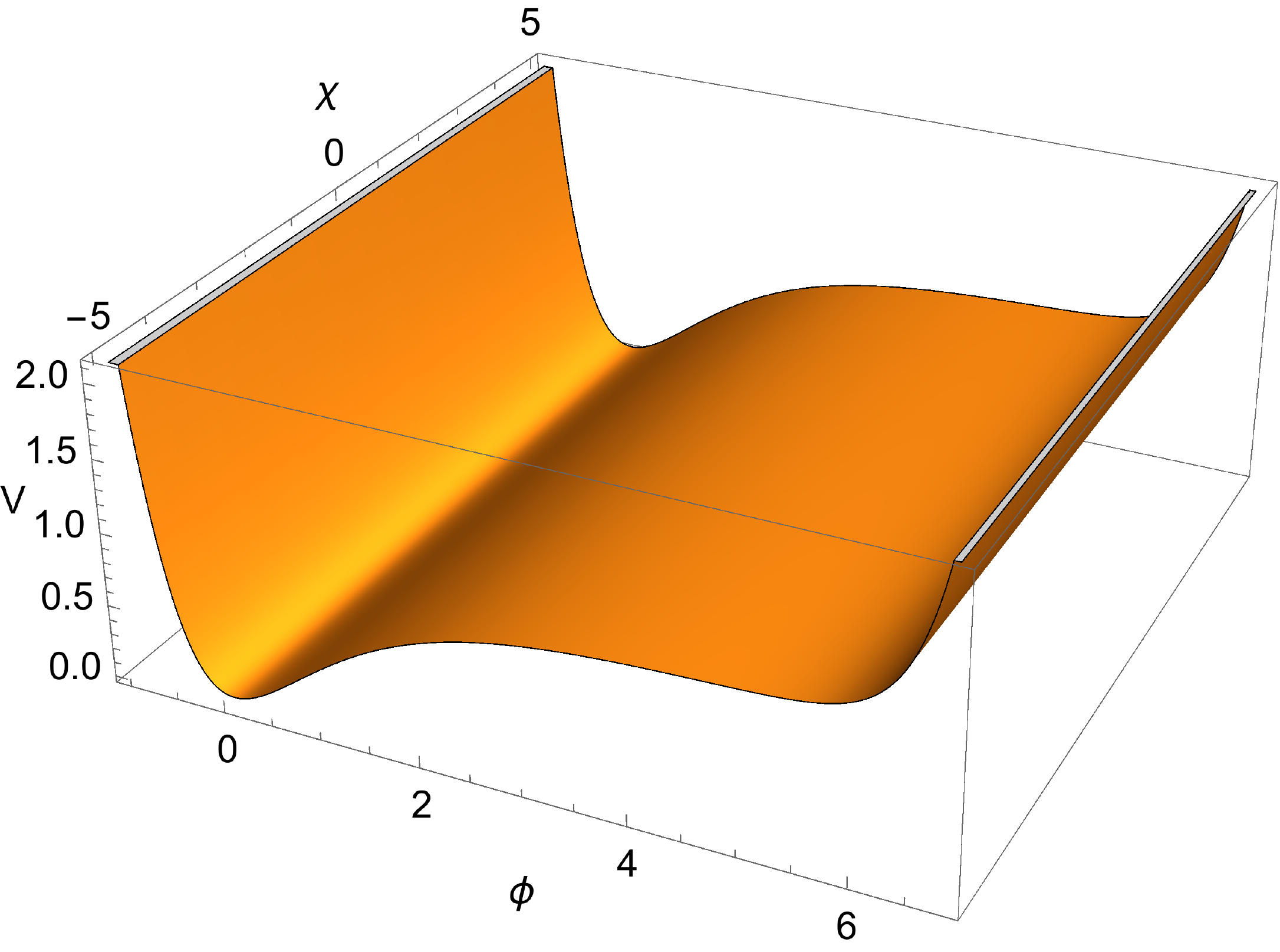}
\end{center}
\caption{\footnotesize The scalar potential for $\phi$ and $\chi$ for the deformed model. }
\label{plot4}
\end{figure}
We take the following parameter set
\begin{equation}
\alpha=\frac23,\quad m=1,\quad A=10^{-10},\quad a=1,\quad W_0=-1.
\end{equation}
With these parameters, the condition $V=0$ at the minimum determines the value of $\delta$ as $\delta=1.47\times10^{-8}$. We show the potential in Fig.~\ref{plot4} with this set of parameters. Although the potential becomes steep for sufficiently large $\phi$, there is a plateau region where inflation can take place. The similar bending of the potential occurs in fibre inflation~\cite{Cicoli:2008gp,Kallosh:2017wku} in string theory because of the string loop corrections. As long as the non-perturbative correction is small, the parameter $\delta$ in $\Delta V_0$ is small and the length of plateau region for $\phi$ can be long enough to support inflation for $N>60$.

We finally note that, in general, there might be superpotential coupling between $S$ and $T$ e.g. $W=Be^{-bT}S$, which would cause the similar problem. Even in such case, one can consider a similar deformation of the potential to avoid negative infinite vacuum as discussed in this section.


\begin{thebibliography}{99}
\bibitem{Kallosh:2013yoa} 
  R.~Kallosh, A.~Linde and D.~Roest,
  ``Superconformal Inflationary $\alpha$-Attractors,''
  JHEP {\bf 1311}, 198 (2013)
  [arXiv:1311.0472 [hep-th]].
  
\bibitem{Ade:2015lrj} 
  P.~A.~R.~Ade {\it et al.} [Planck Collaboration],
``Planck 2015 results. XX. Constraints on inflation,''
  Astron.\ Astrophys.\  {\bf 594}, A20 (2016)
  doi:10.1051/0004-6361/201525898
  [arXiv:1502.02114 [astro-ph.CO]].

  
\bibitem{Kallosh:2015zsa} 
  R.~Kallosh and A.~Linde,
``Escher in the Sky,''
  Comptes Rendus Physique {\bf 16}, 914 (2015)
  [arXiv:1503.06785 [hep-th]].




\bibitem{Carrasco:2015uma} 
  J.~J.~M.~Carrasco, R.~Kallosh, A.~Linde and D.~Roest,
  ``Hyperbolic geometry of cosmological attractors,''
  Phys.\ Rev.\ D {\bf 92}, no. 4, 041301 (2015)
  [arXiv:1504.05557 [hep-th]].
  


\bibitem{Ferrara:2016fwe} 
  S.~Ferrara and R.~Kallosh,
  ``Seven-disk manifold, $\alpha$-attractors, and $B$ modes,''
  Phys.\ Rev.\ D {\bf 94}, no. 12, 126015 (2016)
    [arXiv:1610.04163 [hep-th]].


\bibitem{Kallosh:2017ced} 
  R.~Kallosh, A.~Linde, T.~Wrase and Y.~Yamada,
  ``Maximal Supersymmetry and B-Mode Targets,''
  JHEP {\bf 1704}, 144 (2017)
   [arXiv:1704.04829 [hep-th]].
  


\bibitem{Achucarro:2017ing} 
  A.~Ach\'ucarro, R.~Kallosh, A.~Linde, D.~G.~Wang and Y.~Welling,
  ``Universality of multi-field $\alpha$-attractors,''
  arXiv:1711.09478 [hep-th].
  



\bibitem{Kaloper:2008fb} 
  N.~Kaloper and L.~Sorbo,
  ``A Natural Framework for Chaotic Inflation,''
  Phys.\ Rev.\ Lett.\  {\bf 102}, 121301 (2009)
    [arXiv:0811.1989 [hep-th]].


\bibitem{Kaloper:2011jz} 
  N.~Kaloper, A.~Lawrence and L.~Sorbo,
  ``An Ignoble Approach to Large Field Inflation,''
  JCAP {\bf 1103}, 023 (2011)
    [arXiv:1101.0026 [hep-th]].


\bibitem{Dudas:2014pva} 
  E.~Dudas,
  ``Three-form multiplet and Inflation,''
  JHEP {\bf 1412}, 014 (2014)
   [arXiv:1407.5688 [hep-th]].


\bibitem{Green:1984sg} 
  M.~B.~Green and J.~H.~Schwarz,
  ``Anomaly Cancellation in Supersymmetric D=10 Gauge Theory and Superstring Theory,''
  Phys.\ Lett.\  {\bf 149B}, 117 (1984).


\bibitem{LopesCardoso:1991ifk} 
  G.~Lopes Cardoso and B.~A.~Ovrut,
  ``A Green-Schwarz mechanism for D = 4, N=1 supergravity anomalies,''
  Nucl.\ Phys.\ B {\bf 369}, 351 (1992).


\bibitem{McDonough:2016der} 
  E.~McDonough and M.~Scalisi,
  ``Inflation from Nilpotent K\"ahler Corrections,''
  JCAP {\bf 1611}, no. 11, 028 (2016)
  [arXiv:1609.00364 [hep-th]].


\bibitem{Kallosh:2017wnt} 
  R.~Kallosh, A.~Linde, D.~Roest and Y.~Yamada,
  ``$ \overline{D3} $ induced geometric inflation,''
  JHEP {\bf 1707}, 057 (2017)
  [arXiv:1705.09247 [hep-th]].


\bibitem{Rocek:1978nb} 
  M.~Rocek,
  ``Linearizing the Volkov-Akulov Model,''
  Phys.\ Rev.\ Lett.\  {\bf 41}, 451 (1978).


\bibitem{Lindstrom:1979kq} 
  U.~Lindstrom and M.~Rocek,
  ``Constrained Local Superfields,''
  Phys.\ Rev.\ D {\bf 19}, 2300 (1979).


\bibitem{Casalbuoni:1988xh} 
  R.~Casalbuoni, S.~De Curtis, D.~Dominici, F.~Feruglio and R.~Gatto,
  ``Nonlinear Realization of Supersymmetry Algebra From Supersymmetric Constraint,''
  Phys.\ Lett.\ B {\bf 220}, 569 (1989).


\bibitem{Komargodski:2009rz} 
  Z.~Komargodski and N.~Seiberg,
  ``From Linear SUSY to Constrained Superfields,''
  JHEP {\bf 0909}, 066 (2009)
  [arXiv:0907.2441 [hep-th]].


\bibitem{Kuzenko:2010ef} 
  S.~M.~Kuzenko and S.~J.~Tyler,
  ``Relating the Komargodski-Seiberg and Akulov-Volkov actions: Exact nonlinear field redefinition,''
  Phys.\ Lett.\ B {\bf 698}, 319 (2011)
  [arXiv:1009.3298 [hep-th]].


\bibitem{Kallosh:2017wku} 
  R.~Kallosh, A.~Linde, D.~Roest, A.~Westphal and Y.~Yamada,
  ``Fibre Inflation and $\alpha$-attractors,''
  JHEP {\bf 1802}, 117 (2018)
  [arXiv:1707.05830 [hep-th]].


\bibitem{Hasegawa:2017nks} 
  F.~Hasegawa, K.~Nakayama, T.~Terada and Y.~Yamada,
  ``Gravitino problem in inflation driven by inflaton-polonyi K\"ahler coupling,''
  Phys.\ Lett.\ B {\bf 777}, 270 (2018)
   [arXiv:1709.01246 [hep-ph]].


\bibitem{Burgess:1995kp} 
  C.~P.~Burgess, J.-P.~Derendinger, F.~Quevedo and M.~Quiros,
  ``Gaugino condensates and chiral linear duality: An Effective Lagrangian analysis,''
  Phys.\ Lett.\ B {\bf 348}, 428 (1995)
   [hep-th/9501065].


\bibitem{Burgess:1995aa} 
  C.~P.~Burgess, J.~P.~Derendinger, F.~Quevedo and M.~Quiros,
  ``On gaugino condensation with field dependent gauge couplings,''
  Annals Phys.\  {\bf 250}, 193 (1996)
  [hep-th/9505171].


\bibitem{Binetruy:1995hq} 
  P.~Binetruy, M.~K.~Gaillard and T.~R.~Taylor,
  ``Dynamical supersymmetric breaking and the linear multiplet,''
  Nucl.\ Phys.\ B {\bf 455}, 97 (1995)
  [hep-th/9504143].


\bibitem{Kaku:1978nz} 
  M.~Kaku, P.~K.~Townsend and P.~van Nieuwenhuizen,
  ``Properties of Conformal Supergravity,''
  Phys.\ Rev.\ D {\bf 17}, 3179 (1978).


\bibitem{Kaku:1978ea} 
  M.~Kaku and P.~K.~Townsend,
  ``Poincare Supergravity As Broken Superconformal Gravity,''
  Phys.\ Lett.\  {\bf 76B}, 54 (1978).


\bibitem{Townsend:1979ki} 
  P.~K.~Townsend and P.~van Nieuwenhuizen,
  ``Simplifications of Conformal Supergravity,''
  Phys.\ Rev.\ D {\bf 19}, 3166 (1979).


\bibitem{Kugo:1982cu} 
  T.~Kugo and S.~Uehara,
  ``Conformal and Poincare Tensor Calculi in $N=1$ Supergravity,''
  Nucl.\ Phys.\ B {\bf 226}, 49 (1983).


\bibitem{Kugo:1983mv} 
  T.~Kugo and S.~Uehara,
  ``$N=1$ Superconformal Tensor Calculus: Multiplets With External Lorentz Indices and Spinor Derivative Operators,''
  Prog.\ Theor.\ Phys.\  {\bf 73}, 235 (1985).


\bibitem{Freedman:2012zz} 
  D.~Z.~Freedman and A.~Van Proeyen,
  ``Supergravity,''


\bibitem{Gates:1980ay} 
  S.~J.~Gates, Jr.,
  ``Super P Form Gauge Superfields,''
  Nucl.\ Phys.\ B {\bf 184}, 381 (1981).


\bibitem{Binetruy:1996xw} 
  P.~Binetruy, F.~Pillon, G.~Girardi and R.~Grimm,
  ``The Three form multiplet in supergravity,''
  Nucl.\ Phys.\ B {\bf 477}, 175 (1996)
  [hep-th/9603181].


\bibitem{Ovrut:1997ur} 
  B.~A.~Ovrut and D.~Waldram,
  ``Membranes and three form supergravity,''
  Nucl.\ Phys.\ B {\bf 506}, 236 (1997)
  [hep-th/9704045].


\bibitem{Binetruy:2000zx} 
  P.~Binetruy, G.~Girardi and R.~Grimm,
  ``Supergravity couplings: A Geometric formulation,''
  Phys.\ Rept.\  {\bf 343}, 255 (2001)
  [hep-th/0005225].


\bibitem{Groh:2012tf} 
  K.~Groh, J.~Louis and J.~Sommerfeld,
  ``Duality and Couplings of 3-Form-Multiplets in N=1 Supersymmetry,''
  JHEP {\bf 1305}, 001 (2013)
  [arXiv:1212.4639 [hep-th]].


\bibitem{Becker:2016xgv} 
  K.~Becker, M.~Becker, W.~D.~Linch and D.~Robbins,
  ``Abelian tensor hierarchy in 4D, N = 1 superspace,''
  JHEP {\bf 1603}, 052 (2016)
  [arXiv:1601.03066 [hep-th]].


\bibitem{Becker:2016rku} 
  K.~Becker, M.~Becker, W.~D.~Linch and D.~Robbins,
  ``Chern-Simons actions and their gaugings in 4D, $N =$ 1 superspace,''
  JHEP {\bf 1606}, 097 (2016)
  [arXiv:1603.07362 [hep-th]].


\bibitem{Aoki:2016rfz} 
  S.~Aoki, T.~Higaki, Y.~Yamada and R.~Yokokura,
  ``Abelian tensor hierarchy in 4D ${\cal N}=1$ conformal supergravity,''
  JHEP {\bf 1609}, 148 (2016)
  [arXiv:1606.04448 [hep-th]].


\bibitem{Yokokura:2016xcf} 
  R.~Yokokura,
  ``Abelian tensor hierarchy and Chern-Simons actions in 4D $\mathcal N=1$ conformal supergravity,''
  JHEP {\bf 1612}, 092 (2016)
  [arXiv:1609.01111 [hep-th]].


\bibitem{Becker:2017njd} 
  K.~Becker, M.~Becker, W.~D.~Linch, III, S.~Randall and D.~Robbins,
  ``All Chern-Simons Invariants of 4D, N = 1 Gauged Superform Hierarchies,''
  JHEP {\bf 1704}, 103 (2017)
  [arXiv:1702.00799 [hep-th]].


\bibitem{Becker:2017zwe} 
  K.~Becker, M.~Becker, D.~Butter, S.~Guha, W.~D.~Linch and D.~Robbins,
  ``Eleven-dimensional supergravity in 4D, $N = 1$ superspace,''
  JHEP {\bf 1711}, 199 (2017)
  [arXiv:1709.07024 [hep-th]].


\bibitem{Farakos:2017jme} 
  F.~Farakos, S.~Lanza, L.~Martucci and D.~Sorokin,
  ``Three-forms in Supergravity and Flux Compactifications,''
  Eur.\ Phys.\ J.\ C {\bf 77}, no. 9, 602 (2017)
  [arXiv:1706.09422 [hep-th]].

\bibitem{Ema:2016ops} 
  Y.~Ema, K.~Hamaguchi, T.~Moroi and K.~Nakayama,
  ``Flaxion: a minimal extension to solve puzzles in the standard model,''
  JHEP {\bf 1701}, 096 (2017)
  [arXiv:1612.05492 [hep-ph]].
  
  \bibitem{ours}
A.~Linde, Y.~Yamada, {\it work in progress}.

\bibitem{Linder:2015qxa} 
  E.~V.~Linder,
  ``Dark Energy from $\alpha$-Attractors,''
  Phys.\ Rev.\ D {\bf 91}, no. 12, 123012 (2015)
  [arXiv:1505.00815 [astro-ph.CO]].


\bibitem{Dimopoulos:2017zvq} 
  K.~Dimopoulos and C.~Owen,
  ``Quintessential Inflation with $\alpha$-attractors,''
  JCAP {\bf 1706}, no. 06, 027 (2017)
  [arXiv:1703.00305 [gr-qc]].


\bibitem{Mishra:2017ehw} 
  S.~S.~Mishra, V.~Sahni and Y.~Shtanov,
  ``Sourcing Dark Matter and Dark Energy from $\alpha$-attractors,''
  JCAP {\bf 1706}, no. 06, 045 (2017)
  [arXiv:1703.03295 [gr-qc]].


\bibitem{Dimopoulos:2017tud} 
  K.~Dimopoulos and C.~Owen,
  ``Instant Preheating in Quintessential Inflation with $\alpha$-Attractors,''
  arXiv:1712.01760 [astro-ph.CO].


\bibitem{Akrami:2017cir} 
  Y.~Akrami, R.~Kallosh, A.~Linde and V.~Vardanyan,
  ``Dark energy, $\alpha$-attractors, and large-scale structure surveys,''
  arXiv:1712.09693 [hep-th].


\bibitem{Brown:2017osf} 
  A.~R.~Brown,
  ``Hyperinflation,''
  arXiv:1705.03023 [hep-th].
\bibitem{Mizuno:2017idt} 
  S.~Mizuno and S.~Mukohyama,
  ``Primordial perturbations from inflation with a hyperbolic field-space,''
  Phys.\ Rev.\ D {\bf 96}, no. 10, 103533 (2017)
    [arXiv:1707.05125 [hep-th]].
\bibitem{Cicoli:2008gp} 
  M.~Cicoli, C.~P.~Burgess and F.~Quevedo,
  JCAP {\bf 0903}, 013 (2009)
  doi:10.1088/1475-7516/2009/03/013
  [arXiv:0808.0691 [hep-th]].
\end{thebibliography}
\end{document}